\documentclass[12pt]{article}
\usepackage{amsfonts}
\usepackage{amssymb}
\usepackage{graphics,psboxit,amsmath}

\def\hybrid{\topmargin -20pt    \oddsidemargin 0pt
        \headheight 0pt \headsep 0pt
        \textwidth 6.35in       
        \textheight 9.25in       
        \marginparwidth .875in
        \parskip 5pt plus 1pt   \jot = 1.5ex}

\hybrid

\def\baselinestretch{1.2}

\catcode`\@=11

\def\marginnote#1{}
\newcount\hour
\newcount\minute
\newtoks\amorpm
\hour=\time\divide\hour by60
\minute=\time{\multiply\hour by60 \global\advance\minute by-\hour}
\edef\standardtime{{\ifnum\hour<12 \global\amorpm={am}%
        \else\global\amorpm={pm}\advance\hour by-12 \fi
        \ifnum\hour=0 \hour=12 \fi
        \number\hour:\ifnum\minute<10 0\fi\number\minute\the\amorpm}}
\edef\militarytime{\number\hour:\ifnum\minute<10 0\fi\number\minute}

\def\draftlabel#1{{\@bsphack\if@filesw {\let\thepage\relax
   \xdef\@gtempa{\write\@auxout{\string
      \newlabel{#1}{{\@currentlabel}{\thepage}}}}}\@gtempa
   \if@nobreak \ifvmode\nobreak\fi\fi\fi\@esphack}
        \gdef\@eqnlabel{#1}}
\def\@eqnlabel{}
\def\@vacuum{}
\def\draftmarginnote#1{\marginpar{\raggedright\scriptsize\tt#1}}

\def\draft{\oddsidemargin -.5truein
        \def\@oddfoot{\sl preliminary draft \hfil
        \rm\thepage\hfil\sl\today\quad\militarytime}
        \let\@evenfoot\@oddfoot \overfullrule 3pt
        \let\label=\draftlabel
        \let\marginnote=\draftmarginnote
   \def\@eqnnum{(\theequation)\rlap{\kern\marginparsep\tt\@eqnlabel}%
\global\let\@eqnlabel\@vacuum}  }

\def\preprint{\twocolumn\sloppy\flushbottom\parindent 2em
        \leftmargini 2em\leftmarginv .5em\leftmarginvi .5em
        \oddsidemargin -.5in    \evensidemargin -.5in
        \columnsep .4in \footheight 0pt
        \textwidth 10.in        \topmargin  -.4in
        \headheight 12pt \topskip .4in
        \textheight 6.9in \footskip 0pt
        \def\@oddhead{\thepage\hfil\addtocounter{page}{1}\thepage}
        \let\@evenhead\@oddhead \def\@oddfoot{} \def\@evenfoot{} }

\def\numberbysection{\@addtoreset{equation}{section}
        \def\theequation{\thesection.\arabic{equation}}}

\def\underline#1{\relax\ifmmode\@@underline#1\else
        $\@@underline{\hbox{#1}}$\relax\fi}

\def\titlepage{\@restonecolfalse\if@twocolumn\@restonecoltrue\onecolumn
     \else \newpage \fi \thispagestyle{empty}\c@page\z@
        \def\thefootnote{\fnsymbol{footnote}} }

\def\endtitlepage{\if@restonecol\twocolumn \else \newpage \fi
        \def\thefootnote{\arabic{footnote}}
        \setcounter{footnote}{0}}  

\catcode`@=12
\relax

\def\figcap{\section*{Figure Captions\markboth
        {FIGURECAPTIONS}{FIGURECAPTIONS}}\list
        {Figure \arabic{enumi}:\hfill}{\settowidth\labelwidth{Figure
999:}
        \leftmargin\labelwidth
        \advance\leftmargin\labelsep\usecounter{enumi}}}
 \relax
\def\tablecap{\section*{Table Captions\markboth
        {TABLECAPTIONS}{TABLECAPTIONS}}\list
        {Table \arabic{enumi}:\hfill}{\settowidth\labelwidth{Table
999:}
        \leftmargin\labelwidth
        \advance\leftmargin\labelsep\usecounter{enumi}}}
 \relax
\def\reflist{\section*{References\markboth
        {REFLIST}{REFLIST}}\list
        {[\arabic{enumi}]\hfill}{\settowidth\labelwidth{[999]}
        \leftmargin\labelwidth
        \advance\leftmargin\labelsep\usecounter{enumi}}}
 \relax
\makeatletter
\newcounter{pubctr}
\def\publist{\@ifnextchar[{\@publist}{\@@publist}}
\def\@publist[#1]{\list
        {[\arabic{pubctr}]\hfill}{\settowidth\labelwidth{[999]}
        \leftmargin\labelwidth
        \advance\leftmargin\labelsep
        \@nmbrlisttrue\def\@listctr{pubctr}
        \setcounter{pubctr}{#1}\addtocounter{pubctr}{-1}}}
\def\@@publist{\list
        {[\arabic{pubctr}]\hfill}{\settowidth\labelwidth{[999]}
        \leftmargin\labelwidth
        \advance\leftmargin\labelsep
        \@nmbrlisttrue\def\@listctr{pubctr}}}
 \relax
\makeatother
\newskip\humongous \humongous=0pt plus 1000pt minus 1000pt

\newif\ifdtup

\relax

\def\be{\begin{equation}}
\def\ee{\end{equation}}
\def\ba{\begin{eqnarray}}
\def\ea{\end{eqnarray}}

\def\no{\noindent}

\def\IR{\relax{\rm I\kern-.18em R}}
\def\II{\relax{\rm 1\kern-.35em1}}

\renewcommand{\theequation}{\thesection.\arabic{equation}}
\csname @addtoreset\endcsname{equation}{section}

\def\IR{\relax{\rm I\kern-.18em R}}
\def\inv{^{\raise.15ex\hbox{${\scriptscriptstyle -}$}\kern-.05em 1}}

\begin{document}

\begin{titlepage}
\begin{center}

\hfill IFT-UAM/CSIC-07-61\\
\vskip -.1 cm
\hfill NI-07079
\vskip -.1 cm
\hfill arXiv:0711.3404\\

\vskip .5in

{\LARGE The Ising model and planar ${\cal N}=4$ Yang-Mills}
\vskip 0.4in

{\bf C\'esar G\'omez,}\phantom{x}{\bf Johan Gunnesson}\phantom{x}and\phantom{x}
{\bf Rafael Hern\'andez} 
\vskip 0.1in

Departamento de F\'{\i}sica Te\'orica C-XI
and Instituto de F\'{\i}sica Te\'orica UAM-CSIC, C-XVI \\
Universidad Aut\'onoma de Madrid,
Cantoblanco, 28049 Madrid, Spain\\
{\footnotesize{\tt cesar.gomez@uam.es, johan.gunnesson@uam.es, rafael.hernandez@cern.ch}}

\end{center}

\vskip .4in

\centerline{\bf Abstract}
\vskip .1in
\no
The scattering-matrix for planar Yang-Mills with ${\cal N}=4$ supersymmetry relies on the 
assumption that integrability holds to all orders in perturbation theory. In this note 
we define a map from the spectral variables $x^{\pm}$, parameterizing the long-range 
magnon momenta, to couplings in a two-dimensional Ising model. Under this map 
integrability of planar ${\cal N}=4$ Yang-Mills becomes equivalent to the 
Yang-Baxter equation for the two-dimensional Ising model, and the long-range 
variables $x^{\pm}$ translate into the entries of the Ising transfer matrices. 
We explore the Ising correlation length which equals the inverse magnon 
momentum in the small momentum limit. The critical regime is thus reached 
for vanishing magnon momentum. We also discuss the meaning of the Kramers-Wannier 
duality transformation on the gauge theory, together with that of the Ising model 
critical points.
\noindent

\vskip .4in
\noindent

\end{titlepage}
\vfill
\eject

\def\baselinestretch{1.2}


\baselineskip 20pt


\section{Introduction}

\no
During the last years our understanding of the AdS/CFT correspondence 
has benefited greatly from its apparent integrability. The identification of the one-loop planar 
dilatation operator for Yang-Mills with ${\cal N}=4$ supersymmetry with the hamiltonian of 
an integrable spin chain \cite{MZ,complete}, enabled the use of Bethe ansatz techniques to 
compute anomalous dimensions for large composite gauge-invariant operators. In the 
$su(2)$ sector the integrable system reduces to the $XXX_{1/2}$ Heisenberg spin chain, and 
the dilatation operator can thus be diagonalized by the Bethe ansatz, giving the allowed 
set of magnon momenta $\{p_j\}$ as solutions to
\begin{equation}
e^{ip_jL} = \prod _{i\neq j}^M S(p_i,\, p_j) \ ,
\label{eq:bethe}
\end{equation}
where the scattering-matrix takes the form
\begin{equation}
S(p_i,\, p_j) = \frac{u(p_i)-u(p_j)+i}{u(p_i)-u(p_j)-i} \ ,
\label{eq:Sheisenberg}
\end{equation}
with $u$, the spin chain rapidity, given by
\begin{equation}
u(p) = \frac{1}{2}\cot \frac{p}{2} \ .
\label{eq:uheisenberg}
\end{equation}
Assuming integrability holds to all orders in perturbation theory, a long-range 
Bethe ansatz for asymptotically long spin chains was later on conjectured in 
\cite{BDS}. The S-matrix in the long-range Bethe ansatz takes the same form as in 
\eqref{eq:Sheisenberg}, but the rapidity \eqref{eq:uheisenberg} is replaced by
\begin{equation}
u(p) = \frac{1}{2}\cot \frac{p}{2}\sqrt{1 + g^2 \sin ^2 \frac{p}{2}} \ ,
\label{eq:uall-loop}
\end{equation}
where $g = \frac{\sqrt{\lambda}}{\pi}$, and where $\lambda \equiv g^2_{YM}N$ is 't Hooft's 
coupling constant. \footnote{Note that $g$ is rescaled, as compared to earlier conventions, 
and coincides with the $\gamma$ of \cite{zaremboworldsheet}.}
This conjecture was subsequently extended to other sectors \cite{StaudacherS,longrange}. 
In doing so, introducing a set of spectral variables $x^+$ and $x^-$ proved convenient. 
They are defined through the relations
\begin{equation}
e^{ip}= \frac{x^+}{x^-} \ ,
\label{eq:eip}
\end{equation}
and
\begin{equation}
x^+ + \frac{1}{x^+} - x^- - \frac{1}{x^-} = \frac{4i}{g} \ .
\label{eq:closure}
\end{equation}
After a rescaling $u \rightarrow \frac{4 u}{g}$, the long-range spin chain rapidity \eqref{eq:uall-loop} 
takes a quite simple form in terms of $x^\pm$,
\begin{equation}
u= \frac{1}{2}\left(x^+ + \frac{1}{x^+} + x^- + \frac{1}{x^-} \right) \ . 
\label{eq:u}
\end{equation}

In \cite{BeisertS} and \cite{BeisertS2}, the long-range S-matrix for planar ${\cal N}=4$ 
Yang-Mills was then constructed algebraically \footnote{Algebraic considerations fix 
the S-matrix up to a global dressing phase factor. The dressing phase is 
constrained by the integrable structure of semiclassical strings \cite{AFS}, or by the 
first quantum correction \cite{HL} (see also \cite{Gromov}). A solution to the algebraic condition that 
crossing symmetry imposes on the dressing factor \cite{Janik} allowed an all-order 
strong-coupling expansion \cite{BHL}, that lead to agreement \cite{BES} with a 
perturbative computation in the weak-coupling regime. To date there is however no general 
symmetry prescription to fix or determine unambiguously \cite{nonpert} the structure of the 
dressing phase factor.}
by demanding invariance of the S-matrix under a centrally-extended $su(2|2)$ algebra. This algebraic construction 
is performed as follows. First of all one should identify magnons with $(2|2)$ irreducible representations of the 
centrally extended algebra. These irreps are parameterized by the eigenvalues of the central elements. Secondly, 
the action of the algebra must be lifted to two-magnon states. This introduces a co-multiplication rule 
that by consistency should be an algebra homomorphism. For a classical algebra this co-multiplication, 
or composition rule, takes the standard form
$\Delta \mathfrak{J} = \mathfrak{J} \otimes \II + \II \otimes \mathfrak{J}$, 
with $\mathfrak{J}$ any algebra generator.
Finally the two-magnon S-matrix is determined by imposing
\begin{equation}
S \, \Delta _{12}(\mathfrak{J}) = \Delta _{21}(\mathfrak{J}) \, S \ ,
\label{eq:Smatrixconstruction}
\end{equation}
where $\Delta _{12}(\mathfrak{J})$ means the action of $\mathfrak{J}$ on two incoming magnons, 
labelled $1$ and $2$. In order to have a non-trivial S-matrix, different from just a permutation, 
we need an asymmetric co-multiplication rule. This is indeed the typical situation in quantum deformed algebras. 
The crucial step in Beisert's algebraic construction of the long-range S-matrix was to define a 
non-symmetric co-multiplication for the generators of the centrally-extended $su(2|2)$ algebra by introducing 
a new generator, the magnon momentum \cite{BeisertS}. The asymmetric 
co-multiplication for the central elements is given by \cite{Hopf} (see also \cite{Plefka}-\cite{BeisertYangian})
\begin{align}
\Delta \mathfrak{P} & = \mathfrak{P} \otimes e^{i\hat{p}} + \II \otimes \mathfrak{P} \ , \\
\Delta \mathfrak{K} & = \mathfrak{K} \otimes e^{-i\hat{p}} + \II \otimes \mathfrak{K} \ ,
\end{align}
while the co-products for the rest of the generators of the algebra are taken to be compatible with 
those of the central charges \cite{BeisertYangian,Young}. 
In principle these co-multiplication rules will define a Hopf algebra structure, 
with generators those in the centrally extended $su(2|2)$ algebra, together with the magnon momentum operator.
Taking now into acccount that central elements commute with the S-matrix, condition 
\eqref{eq:Smatrixconstruction} leads to the constraint $\Delta _{12}(\mathfrak{L}) = \Delta _{21}(\mathfrak{L})$, 
with $\mathfrak{L}$ any central element of the algebra. Using the asymmetric co-multiplications 
defined above, these relations allow us to relate the labels of the magnon irreps, {\em i.e.} the eigenvalues of the central 
elements, to the magnon momentum. In this setup, once we introduce the $x^{\pm}$ variables through 
$\frac{x^+}{x^-} = e^{ip}$, we get Beisert's parametrization of the irrep in terms of the generalized 
rapidities \cite{BeisertS}. It is important to keep in mind that hidden in the parametrization of the magnon irreps in 
terms of the $x^\pm$ variables there is a non-symmetric co-multiplication. This co-multiplication, together with 
the introduction of the extra momentum generator, are two ingredients that by no means are contained 
in the classical centrally extended $su(2|2)$ algebra, encoding the {\em classical} symmetries of the problem. 
  
The form of the co-multiplication already provides some hints on the underlying physics. The quantity 
that is playing the role of a measure for the deformation of the algebra is the magnon momentum. In fact, for zero magnon 
momentum the co-multiplication becomes classical, and we should expect the S-matrix to be simply a permutation. 
If the S-matrix is expanded in the incoming magnon momenta $p_1$ and $p_2$, it takes the form
\begin{equation}
{\cal S} = \II + S_{1,1}p_1 + S_{1,2}p_2 + \cdots
\label{eq:Spexpansion}
\end{equation}
The point is therefore that the $x^+$ and $x^-$ variables describe the departure of the S-matrix 
from triviality, while the classical algebra determines the precise form of the entries of the S-matrix 
in terms of the $x^{\pm}$. But there is yet another motivation to clarify the precise meaning of the $x^{\pm}$ 
spectral variables. The long-range Bethe ansatz is asymptotic, and its validity is in fact limited by wrapping 
effects (see for instance \cite{Lipatov}). If one wishes to extend the integrable spin chain to non-asymptotically 
long chains it is crucial to clarify the meaning of the $x^{\pm}$ variables.

The purpose of this note is to show that there is a way to map the $x^\pm$ variables 
into Ising model couplings $K$ and $L$. In this way a natural interpretation will arise for 
the long-range spin chain rapidity $u$ in terms of Ising model quantities. Under this correspondance, 
the Yang-Baxter equations for the Ising model are completely equivalent to the closure relation \eqref{eq:closure}, 
which we will prove to be equivalent (not just implying) to the $su(2|2)$ spin chain Yang-Baxter equations. 
Furthermore, we will show that the Ising model correlation length seems to be related to the deformation parameter 
of the Hopf algebra of the theory. There is also a possibility, as we will motivate, that 
the Kramers-Wannier duality of the model could play a role in the full, supposedly integrable, 
planar $\mathcal{N}=4$ Yang-Mills theory.


\section{The map to the Ising model }

\no
In this section we will exhibit how the dynamics of planar ${\cal N}=4$ Yang-Mills can be 
mapped to the two-dimensional Ising model. The Ising model 
on a square lattice is defined in terms of horizontal and vertical 
couplings $J$ and $J'$, the temperature $T$ and Boltzmann's constant $k_B$. Following Baxter \cite{Baxter}, 
we will define new couplings, $K$ and $L$, by $K = J/k_BT$ and $L = J'/k_BT$. 
The Ising model partition function is then given by
\begin{equation}
Z = \sum \prod _{(i,\, j)_H} \prod _{(k,\, l)_V} e^{K \sigma_i \cdot \sigma_j + L \sigma_k \cdot \sigma_l} \ ,
\label{eq:Isingpart}
\end{equation}
where $\sigma _i = \pm 1$ is the spin sitting at site $i$, the sum is taken over all spin 
configurations, and $\{(i,\, j)_H \}$ stands for the set of sites, adjacent in the horizontal direction, 
while $\{(i,\, j)_V \}$ is defined analogously for the vertical direction.

In this note we will propose a map from the $x^{\pm}$ variables, describing a magnon in the long-range 
spin chain of \cite{BeisertS2}, 
to Ising model couplings through the relation \footnote{This parameterization of $x^\pm$ is similar 
to the one employed in \cite{Kostov} in terms of $p$ and $\beta$.}
\begin{equation}
x^{\pm} = e^{-2L}e^{\pm 2K} \ .
\label{eq:Isingmap}
\end{equation}
The main theme of this work will be the study of this map and see what light it sheds on the long-range spin chain 
for ${\cal N}=4$ Yang-Mills. 
To begin with, it is immediate to relate the Ising couplings to more familiar quantities appearing in the spin chain. 
Since $e^{ip} \equiv \frac{x^+}{x^-}$, the coupling $K$ is simply $\frac{ip}{4}$. Furthermore, from \cite{BeisertS2}, 
the eigenvalue $C$ of the central charge $\mathfrak{C}$, usually interpreted as the magnon energy, 
is given by $C = \frac{1}{2}\frac{1 + 1/x^+x^-}{1 - 1/x^+x^-}$, which, using the map 
\eqref{eq:Isingmap}, just becomes $-\frac{1}{2}\coth 2L$. In conclusion,
\be
e^{ip} = e^{4K}, \quad C = -\frac{1}{2}\coth 2L \ . 
\label{eq:CL}
\ee
It should also be noted that the two possible solutions of \eqref{eq:closure}, in the limit 
$g \rightarrow \infty$, normally given as $x^+ = x^-$ or $x^+ = 1/x^-$, now correspond to letting $K \rightarrow 0$ 
or $L \rightarrow 0$, respectively. 
However, a more important consequence of \eqref{eq:Isingmap} is that spectral variables $x^\pm$ can 
be given a direct interpretation. To do so, we will study the Ising model transfer matrices.


\subsection{Ising transfer matrices}

\no 
A standard way of calculating the partition function \eqref{eq:Isingpart} is by introducing transfer 
matrices $V$ and $W$. Rotating the lattice by $45^\circ $, these can be described graphically as in 
figures 1 and 2. From the spin configurations $a_1,\, a_2,\, \ldots,\, a_n$ and 
$b_1,\, b_2,\, \ldots ,\, b_n$, one obtains the matrix elements of the transfer matrices by multiplying 
the Boltzmann weights corresponding to the lines connecting each of the sites. The relevant Boltzmann weights 
are $e^{\pm L}$ for lines marked $L$, and $e^{\pm K}$ for lines marked $K$, with the plus sign for lines 
connecting spins of the same type, and the minus sign when the adjacent spins are of opposite type.
\setlength{\unitlength}{2368sp}%
\begingroup\makeatletter\ifx\SetFigFont\undefined
\def\x#1#2#3#4#5#6#7\relax{\def\x{#1#2#3#4#5#6}}%
\expandafter\x\fmtname xxxxxx\relax \def\y{splain}%
\ifx\x\y   
\gdef\SetFigFont#1#2#3{%
  \ifnum #1<17\tiny\else \ifnum #1<20\small\else
  \ifnum #1<24\normalsize\else \ifnum #1<29\large\else
  \ifnum #1<34\Large\else \ifnum #1<41\LARGE\else
     \huge\fi\fi\fi\fi\fi\fi
  \csname #3\endcsname}%
\else
\gdef\SetFigFont#1#2#3{\begingroup
  \count@#1\relax \ifnum 25<\count@\count@25\fi
  \def\x{\endgroup\@setsize\SetFigFont{#2pt}}%
  \expandafter\x
    \csname \romannumeral\the\count@ pt\expandafter\endcsname
    \csname @\romannumeral\the\count@ pt\endcsname
  \csname #3\endcsname}%
\fi
\fi\endgroup
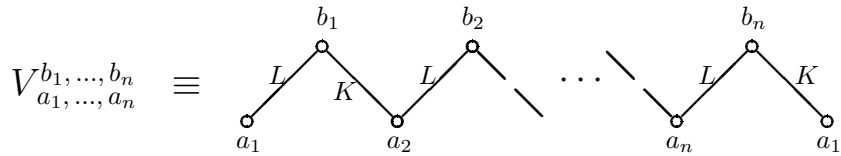
\begin{figure}
\begin{center}
\begin{picture}(8628,1679)(1576,-2697)
\put(8775,-1917){\makebox(0,0)[lb]{\smash{\SetFigFont{10}{12.0}{rm}$L$}}}
\thicklines
\put(6413,-1523){\oval(112,112)}
\put(5626,-2311){\oval(112,112)}
\put(6413,-1523){\oval(112,112)}
\put(5626,-2311){\oval(112,112)}
\put(4838,-1523){\oval(112,112)}
\put(8550,-2310){\oval(112,112)}
\put(9338,-1523){\oval(112,112)}
\put(10125,-2310){\oval(112,112)}
\put(4782,-1579){\line(-1,-1){675}}
\put(5682,-2254){\line( 1, 1){675}}
\put(4894,-1579){\line( 1,-1){675}}
\put(6469,-1579){\line( 1,-1){282}}
\put(6864,-1974){\line( 1,-1){280}}
\put(7820,-1580){\line( 1,-1){281}}
\put(8213,-1973){\line( 1,-1){281}}
\put(8607,-2254){\line( 1, 1){675}}
\put(9394,-1579){\line( 1,-1){675}}
\put(1576,-2030){\makebox(0,0)[lb]{\smash{\SetFigFont{14}{16.8}{rm}$V_{\, a_{1},\, \ldots,\, a_{n}}^{b_{1},\, \ldots,\, b_{n}}$}}}
\put(3262,-2030){\makebox(0,0)[lb]{\smash{\SetFigFont{14}{16.8}{rm}$\equiv$}}}
\put(4275,-1917){\makebox(0,0)[lb]{\smash{\SetFigFont{10}{12.0}{rm}$L$}}}
\put(5850,-1917){\makebox(0,0)[lb]{\smash{\SetFigFont{10}{12.0}{rm}$L$}}}
\put(4951,-2085){\makebox(0,0)[lb]{\smash{\SetFigFont{10}{12.0}{rm}$K$}}}
\put(7313,-1973){\makebox(0,0)[lb]{\smash{\SetFigFont{14}{16.8}{rm}$\cdots $}}}
\put(3938,-2592){\makebox(0,0)[lb]{\smash{\SetFigFont{10}{12.0}{rm}$a_{1}$}}}
\put(10013,-2592){\makebox(0,0)[lb]{\smash{\SetFigFont{10}{12.0}{rm}$a_{1}$}}}
\put(5513,-2592){\makebox(0,0)[lb]{\smash{\SetFigFont{10}{12.0}{rm}$a_{2}$}}}
\put(8438,-2592){\makebox(0,0)[lb]{\smash{\SetFigFont{10}{12.0}{rm}$a_{n}$}}}
\put(4781,-1298){\makebox(0,0)[lb]{\smash{\SetFigFont{10}{12.0}{rm}$b_{1}$}}}
\put(6300,-1298){\makebox(0,0)[lb]{\smash{\SetFigFont{10}{12.0}{rm}$b_{2}$}}}
\put(9225,-1298){\makebox(0,0)[lb]{\smash{\SetFigFont{10}{12.0}{rm}$b_{n}$}}}
\put(9788,-1917){\makebox(0,0)[lb]{\smash{\SetFigFont{10}{12.0}{rm}$K$}}}
\put(4051,-2311){\oval(112,112)}
\end{picture}
\end{center}
\caption{Graphical representation of the transfer matrix V.}
\end{figure}
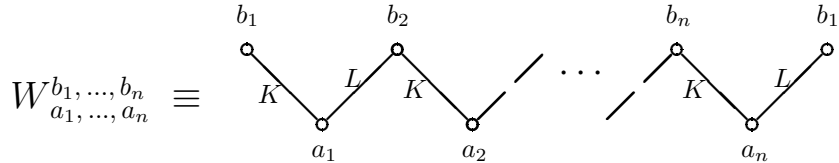
\begin{figure}
\begin{center}
\begin{picture}(8628,1791)(1576,-2640)
\put(9562,-1804){\makebox(0,0)[lb]{\smash{\SetFigFont{10}{12.0}{rm}$L$}}}
\thicklines
\put(9338,-2199){\oval(112,112)}
\put(8550,-1412){\oval(112,112)}
\put(4838,-2199){\oval(112,112)}
\put(5626,-1411){\oval(112,112)}
\put(6413,-2199){\oval(112,112)}
\put(5626,-1411){\oval(112,112)}
\put(6413,-2199){\oval(112,112)}
\put(4051,-1411){\oval(112,112)}
\put(9394,-2143){\line( 1, 1){675}}
\put(8607,-1468){\line( 1,-1){675}}
\put(8213,-1749){\line( 1, 1){281}}
\put(7820,-2142){\line( 1, 1){281}}
\put(6864,-1748){\line( 1, 1){280}}
\put(6469,-2143){\line( 1, 1){282}}
\put(4894,-2143){\line( 1, 1){675}}
\put(5682,-1468){\line( 1,-1){675}}
\put(4782,-2143){\line(-1, 1){675}}
\put(7313,-1749){\makebox(0,0)[lb]{\smash{\SetFigFont{14}{16.8}{rm}$\cdots $}}}
\put(10012,-1129){\makebox(0,0)[lb]{\smash{\SetFigFont{10}{12.0}{rm}$b_{1}$}}}
\put(5682,-1916){\makebox(0,0)[lb]{\smash{\SetFigFont{10}{12.0}{rm}$K$}}}
\put(1576,-2030){\makebox(0,0)[lb]{\smash{\SetFigFont{14}{16.8}{rm}$W_{\, a_{1},\, \ldots,\, a_{n}}^{b_{1},\, \ldots,\, b_{n}}$}}}
\put(3262,-2030){\makebox(0,0)[lb]{\smash{\SetFigFont{14}{16.8}{rm}$\equiv$}}}
\put(5062,-1804){\makebox(0,0)[lb]{\smash{\SetFigFont{10}{12.0}{rm}$L$}}}
\put(4163,-1972){\makebox(0,0)[lb]{\smash{\SetFigFont{10}{12.0}{rm}$K$}}}
\put(4725,-2535){\makebox(0,0)[lb]{\smash{\SetFigFont{10}{12.0}{rm}$a_{1}$}}}
\put(6300,-2535){\makebox(0,0)[lb]{\smash{\SetFigFont{10}{12.0}{rm}$a_{2}$}}}
\put(9225,-2535){\makebox(0,0)[lb]{\smash{\SetFigFont{10}{12.0}{rm}$a_{n}$}}}
\put(3937,-1129){\makebox(0,0)[lb]{\smash{\SetFigFont{10}{12.0}{rm}$b_{1}$}}}
\put(5512,-1129){\makebox(0,0)[lb]{\smash{\SetFigFont{10}{12.0}{rm}$b_{2}$}}}
\put(8437,-1129){\makebox(0,0)[lb]{\smash{\SetFigFont{10}{12.0}{rm}$b_{n}$}}}
\put(8606,-1917){\makebox(0,0)[lb]{\smash{\SetFigFont{10}{12.0}{rm}$K$}}}
\put(10125,-1412){\oval(112,112)}
\end{picture}
\end{center}
\caption{Graphical representation of the transfer matrix W.}
\end{figure}
When the total number of (diagonal) rows $m$ is pair, the partition function is given by 
\begin{equation}
Z = \text{Tr} [(VW)^{m/2}] \ .
\label{eq:Z}
\end{equation}
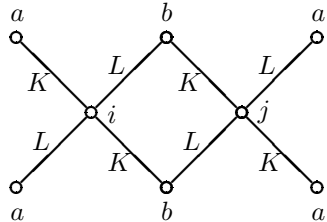
\begin{figure}[h]
\begin{center}
\begin{picture}(3308,2374)(3916,-3458)
\put(6525,-2311){\makebox(0,0)[lb]{\smash{\SetFigFont{10}{12.0}{rm}$j$}}}
\thicklines
\put(5570,-3042){\oval(112,112)}
\put(6357,-2255){\oval(112,112)}
\put(7145,-3042){\oval(112,112)}
\put(5570,-3042){\oval(112,112)}
\put(6357,-2255){\oval(112,112)}
\put(3995,-3042){\oval(112,112)}
\put(4838,-2311){\line( 1,-1){675}}
\put(5626,-2986){\line( 1, 1){675}}
\put(6413,-2311){\line( 1,-1){675}}
\put(4726,-2311){\line(-1,-1){675}}
\put(5570,-1467){\oval(112,112)}
\put(7145,-1467){\oval(112,112)}
\put(5570,-1467){\oval(112,112)}
\put(3995,-1467){\oval(112,112)}
\put(4838,-2198){\line( 1, 1){675}}
\put(5626,-1523){\line( 1,-1){675}}
\put(6413,-2198){\line( 1, 1){675}}
\put(4726,-2198){\line(-1, 1){675}}
\put(7088,-1298){\makebox(0,0)[lb]{\smash{\SetFigFont{10}{12.0}{rm}$a$}}}
\put(3938,-1298){\makebox(0,0)[lb]{\smash{\SetFigFont{10}{12.0}{rm}$a$}}}
\put(7088,-3379){\makebox(0,0)[lb]{\smash{\SetFigFont{10}{12.0}{rm}$a$}}}
\put(5513,-1297){\makebox(0,0)[lb]{\smash{\SetFigFont{10}{12.0}{rm}$b$}}}
\put(6526,-2873){\makebox(0,0)[lb]{\smash{\SetFigFont{10}{12.0}{rm}$K$}}}
\put(5738,-2647){\makebox(0,0)[lb]{\smash{\SetFigFont{10}{12.0}{rm}$L$}}}
\put(6525,-1860){\makebox(0,0)[lb]{\smash{\SetFigFont{10}{12.0}{rm}$L$}}}
\put(4163,-2647){\makebox(0,0)[lb]{\smash{\SetFigFont{10}{12.0}{rm}$L$}}}
\put(4951,-2873){\makebox(0,0)[lb]{\smash{\SetFigFont{10}{12.0}{rm}$K$}}}
\put(4107,-2030){\makebox(0,0)[lb]{\smash{\SetFigFont{10}{12.0}{rm}$K$}}}
\put(4951,-1860){\makebox(0,0)[lb]{\smash{\SetFigFont{10}{12.0}{rm}$L$}}}
\put(5682,-2030){\makebox(0,0)[lb]{\smash{\SetFigFont{10}{12.0}{rm}$K$}}}
\put(3938,-3379){\makebox(0,0)[lb]{\smash{\SetFigFont{10}{12.0}{rm}$a$}}}
\put(5513,-3379){\makebox(0,0)[lb]{\smash{\SetFigFont{10}{12.0}{rm}$b$}}}
\put(4950,-2367){\makebox(0,0)[lb]{\smash{\SetFigFont{10}{12.0}{rm}$i$}}}
\put(4782,-2255){\oval(112,112)}
\end{picture}
\end{center}
\caption{The lattice used in the interpretation of $x^{\pm}$.}
\end{figure}
Let us now consider a small, square lattice with only two rows of two sites each, and 
periodic boundary conditions (see figure 3). 
We can then define $V$ and $W$ transfer matrices, as in the general case. 
The corresponding graphical representation is shown in figures 4 and 5.
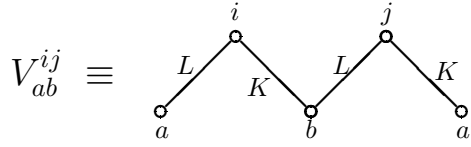
\begin{figure}
\begin{center}
\begin{picture}(4804,1529)(2476,-2671)
\put(2476,-2030){\makebox(0,0)[lb]{\smash{\SetFigFont{14}{16.8}{rm}$V_{ab}^{ij}$}}}
\thicklines
\put(5626,-2311){\oval(112,112)}
\put(6413,-1523){\oval(112,112)}
\put(7201,-2311){\oval(112,112)}
\put(5626,-2311){\oval(112,112)}
\put(6413,-1523){\oval(112,112)}
\put(4051,-2311){\oval(112,112)}
\put(4894,-1579){\line( 1,-1){675}}
\put(5682,-2254){\line( 1, 1){675}}
\put(6469,-1579){\line( 1,-1){675}}
\put(4782,-1579){\line(-1,-1){675}}
\put(4951,-2142){\makebox(0,0)[lb]{\smash{\SetFigFont{10}{12.0}{rm}$K$}}}
\put(5850,-1917){\makebox(0,0)[lb]{\smash{\SetFigFont{10}{12.0}{rm}$L$}}}
\put(4219,-1917){\makebox(0,0)[lb]{\smash{\SetFigFont{10}{12.0}{rm}$L$}}}
\put(6919,-1974){\makebox(0,0)[lb]{\smash{\SetFigFont{10}{12.0}{rm}$K$}}}
\put(3994,-2592){\makebox(0,0)[lb]{\smash{\SetFigFont{10}{12.0}{rm}$a$}}}
\put(7144,-2592){\makebox(0,0)[lb]{\smash{\SetFigFont{10}{12.0}{rm}$a$}}}
\put(5569,-2592){\makebox(0,0)[lb]{\smash{\SetFigFont{10}{12.0}{rm}$b$}}}
\put(4782,-1355){\makebox(0,0)[lb]{\smash{\SetFigFont{10}{12.0}{rm}$i$}}}
\put(6357,-1355){\makebox(0,0)[lb]{\smash{\SetFigFont{10}{12.0}{rm}$j$}}}
\put(3262,-2030){\makebox(0,0)[lb]{\smash{\SetFigFont{14}{16.8}{rm}$\equiv$}}}
\put(4838,-1523){\oval(112,112)}
\end{picture}
\end{center}
\caption{The V transfer matrix for rows of only two sites.}
\end{figure}

\begin{figure}
\begin{center}
\begin{picture}(4748,1586)(2476,-2671)
\put(4107,-2030){\makebox(0,0)[lb]{\smash{\SetFigFont{10}{12.0}{rm}$K$}}}
\thicklines
\put(5570,-1467){\oval(112,112)}
\put(6357,-2254){\oval(112,112)}
\put(7145,-1467){\oval(112,112)}
\put(5570,-1467){\oval(112,112)}
\put(6357,-2254){\oval(112,112)}
\put(3995,-1467){\oval(112,112)}
\put(4838,-2198){\line( 1, 1){675}}
\put(5626,-1523){\line( 1,-1){675}}
\put(6413,-2198){\line( 1, 1){675}}
\put(4726,-2198){\line(-1, 1){675}}
\put(7088,-1298){\makebox(0,0)[lb]{\smash{\SetFigFont{10}{12.0}{rm}$i$}}}
\put(2476,-2030){\makebox(0,0)[lb]{\smash{\SetFigFont{14}{16.8}{rm}$W_{ab}^{ij}$}}}
\put(3262,-2030){\makebox(0,0)[lb]{\smash{\SetFigFont{14}{16.8}{rm}$\equiv$}}}
\put(5513,-1298){\makebox(0,0)[lb]{\smash{\SetFigFont{10}{12.0}{rm}$j$}}}
\put(3938,-1298){\makebox(0,0)[lb]{\smash{\SetFigFont{10}{12.0}{rm}$i$}}}
\put(6301,-2592){\makebox(0,0)[lb]{\smash{\SetFigFont{10}{12.0}{rm}$b$}}}
\put(4726,-2592){\makebox(0,0)[lb]{\smash{\SetFigFont{10}{12.0}{rm}$a$}}}
\put(5682,-2030){\makebox(0,0)[lb]{\smash{\SetFigFont{10}{12.0}{rm}$K$}}}
\put(5007,-1804){\makebox(0,0)[lb]{\smash{\SetFigFont{10}{12.0}{rm}$L$}}}
\put(6806,-2029){\makebox(0,0)[lb]{\smash{\SetFigFont{10}{12.0}{rm}$L$}}}
\put(4782,-2254){\oval(112,112)}
\end{picture}
\end{center}
\caption{The W transfer matrix for rows of only two sites.}
\end{figure}
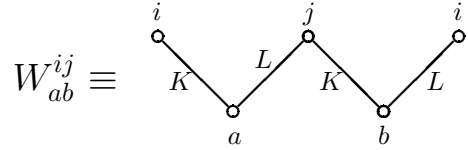
As an example, consider $V$ in the case where $a=j=+$, $b=i=-$. Then, the line connecting $a$ 
and $i$ gives a factor of $e^{-L}$, the line connecting $i$ and $b$ gives $e^K$, the one between 
$b$ and $j$ gives $e^{-L}$ and the line between $j$ and $a$ gives $e^{K}$. Multiplying these four 
factors, we see that $V_{+-}^{-+}=e^{-2L}e^{2K}\equiv x^+$. In this way, all the elements 
of $V$ and $W$ are determined. The result is rather surprising: written in the 
basis $(++,\, +-,\, -+,\, --)$, we find
\begin{equation}
V = 
\left(\begin{array}{cccc}
1/x^- &   1           &   1           &  x^- \\
     1        & 1/x^+ &  x^+          &  1   \\
     1        &  x^+          & 1/x^+ &  1   \\
    x^-       &   1           &   1           &  1/x^- 	
\end{array} \right) \ ,
\label{eq:V}
\end{equation}
and
\begin{equation}
W = 
\left(\begin{array}{cccc}
1/x^-         &   1            &   1      &  x^- \\
     1        &  x^+           &  1/x^+   &  1   \\
     1        &  1/x^+         &  x^+     &  1   \\
    x^-       &   1            &   1      &  1/x^- 	
\end{array} \right) \ .
\label{eq:W}
\end{equation}
We thus see that the generalized rapidities $x^\pm$ are simply the matrix elements of these transfer matrices!
The reader might object that the case where the number $n$ of sites per row is 2 is highly restrictive. 
However, as long as $n$ is even the matrix elements of the corresponding transfer matrices can always be written as
\be
(x^+)^a(x^-)^b,\, \; \text{ for integer $a$ and $b$.}
\label{eq:matrixelement} 
\ee
This is easily seen as follows: 
\begin{enumerate}
\item The matrix elements $V_{\,\, +,\, +, \, \ldots ,\, +}^{+,\, +, \, \ldots ,\, +}$ 
and $W_{\, +,\, +, \, \ldots ,\, +}^{+,\, +, \, \ldots ,\, +}$ can obviously be written in this way, in the form 
$1/ (x^-)^{n/2}$. All matrix elements can then be obtained from these two by flipping some of 
the spins on the upper and lower rows.
\item If a matrix element is of the form \eqref{eq:matrixelement}, then any element 
obtained by flipping a spin $\sigma $ will also be. The site at which $\sigma$ sits is connected to a spin $\rho$ via an 
$L$-line, and to a spin $\tau$ via a $K$-line. If all three spins are equal, flipping $\sigma$ will multiply 
the matrix element by $e^{-2L}e^{-2K} = x^-$. If $\sigma \neq \rho = \tau$, the flip multiplies the element 
by $e^{2L}e^{2K} = 1/x^-$. When $\sigma = \rho \neq \tau$, the multiple is $e^{-2L}e^{2K} = x^+$. And 
if $\sigma = \tau \neq \rho$, one obtains $e^{2L}e^{-2K} = 1/x^+$.
\end{enumerate}
Thus in the general case the $x^{\pm}$ are still natural variables for parameterizing the transfer 
matrices $V$ and $W$. \footnote{The spectral variables $x^{\pm}$ have appeared before in the algebraic 
Bethe ansatz solution of the Hubbard model \cite{Martins1} (see also \cite{Martins2} for a more 
direct relation with the Hubbard model in ${\cal N}=4$ Yang-Mills \cite{Hubbard}).}
  
\subsubsection*{Crossing symmetry}

\no
Let us now address the issue of crossing symmetry at the level of the transfer matrices. 
There is evidence that the S-matrix in the AdS/CFT correspondence exhibits a crossing 
symmetry \cite{Janik}, under which the $x^{\pm}$-variables transform as
\be
\left( x^{\pm}\right)^{\text{cr}} = \frac{1}{x^{\pm}} \ ,
\label{eq:xcrossing}
\ee
where the superscript $^{\text{cr}}$ denotes the crossing transformation. At this point, the reader 
might wonder why we have chosen to study the map \eqref{eq:Isingmap}, relating the $x^\pm$ to the 
Ising model couplings $K$ and $L$, when the map obtained after performing a crossing transformation,
\be
x^{\pm} = e^{2L}e^{\mp 2K},\, 
\ee
should be on equal footing. The fact is that it really does not matter which map we chose, 
because the Ising model is invariant under this transformation. The easiest way to see this 
is by noting that crossing is equivalent to letting 
\be
K,\, L \mapsto -K,\, -L \ ,
\ee 
which leaves the partition function \eqref{eq:Isingpart} invariant. However, if we study instead 
this invariance at the level of the transfer matrices $V$ and $W$, the result turns out to be rather amusing. 
From the graphical representation of $V$ and $W$ in figures 1 and 2, we see that if we flip a spin $b_i$ 
the contribution to the matrices from the attached lines changes from $e^{\pm K}$ and $e^{\pm L}$ to $e^{\mp K}$ 
and $e^{\mp L}$. Changing thus the sign of $K$ and $L$ globally is equivalent to fliping all 
the spins on either the lower or the upper rows of $V$ and $W$. Therefore, denoting the opposite 
of the spin $b_i$ by $\bar{b}_i$ and using collective indeces such as $a \equiv (a_{1},\, \ldots,\, a_{n})$, 
the transfer matrices transform as
\be
\left(V ^{\text{cr}} \right)_{a}^{b} = V_{\, \, a}^{\bar{b}} = 
V_{\, \, \bar{a}}^{b} \ , \;\;\; \left(W ^{\text{cr}} \right)_{a}^{b} = 
W_{\, a}^{\bar{b}} = W_{\, \bar{a}}^{b} \ .
\ee
Then
\be
\left(\left(VW\right)^{\text{cr}} \right)_{a}^{c} = 
\displaystyle \sum_{b}\left(V ^{\text{cr}} \right)_{b}^{c} \left(W ^{\text{cr}} \right)_{a}^{b} 
= \displaystyle \sum_{b}V_{\, \, \bar{b}}^{c} \, W_{\, a}^{\bar{b}} = 
\displaystyle \sum_{\bar{b}}V_{\, \, \bar{b}}^{c} \, W_{\, a}^{\bar{b}} = 
\left(VW \right)_{a}^{c} \ .
\ee
From \eqref{eq:Z} it immediately follows that the partition function is invariant under crossing 
symmetry.


\subsection{Yang-Baxter equation}

We will now take the map from the long-range ${\cal N}=4$ spin chain to the Ising model one step further. 
Let us associate a magnon, given by $x_1^\pm$, with the matrix $V$, and a second magnon, $x_2^\pm$, with $W$. 
Integrability of the Ising model, {\em i.e.} the existence of an infinite number of conserved charges, is encoded 
in the condition
\begin{equation}
V(x_1^\pm)W(x_2^\pm) = V(x_2^\pm)W(x_1^\pm) \ .
\label{eq:Isingintegrability}
\end{equation}
This is true in general, but the connection with the spin chain is clearer for the two-site transfer 
matrices given in \eqref{eq:V} and \eqref{eq:W}. Then, 
\begin{equation}
\left[ V(x_1^\pm) W(x_2^\pm)  \right]^{cd}_{ab} = X(a\, b\, c\, d;\, x_1^\pm,\, x_2^\pm)\cdot X(b\, a\, d\, c;\, x_1^\pm,\, x_2^\pm) \ ,
\label{eq:VWX}
\end{equation}
with
\begin{equation}
\begin{picture}(6041,2373)(-336,-2727)
\put(5007,-1580){\makebox(0,0)[lb]{\smash{\SetFigFont{10}{12.0}{rm}$i$}}}
\thicklines
\put(5626,-2311){\oval(112,112)}
\put(5625,-735){\oval(112,112)}
\put(4051,-736){\oval(112,112)}
\put(5626,-2311){\oval(112,112)}
\put(4051,-2311){\oval(112,112)}
\put(4894,-1579){\line( 1,-1){675}}
\put(4894,-1466){\line( 1, 1){675}}
\put(4106,-791){\line( 1,-1){675}}
\put(4782,-1579){\line(-1,-1){675}}
\put(-336,-1636){\makebox(0,0)[lb]{\smash{\SetFigFont{12}{14}{rm}
$X(a\, b\, c\, d;\, x_1^\pm,\, x_2^\pm) \, \equiv  \;\; \displaystyle \sum _i$}}}
\put(5288,-1917){\makebox(0,0)[lb]{\smash{\SetFigFont{10}{12.0}{rm}$K_1$}}}
\put(4106,-1243){\makebox(0,0)[lb]{\smash{\SetFigFont{10}{12.0}{rm}$K_{2}$}}}
\put(5343,-1242){\makebox(0,0)[lb]{\smash{\SetFigFont{10}{12.0}{rm}$L_{2}$}}}
\put(4106,-1916){\makebox(0,0)[lb]{\smash{\SetFigFont{10}{12.0}{rm}$L_{1}$}}}
\put(5569,-567){\makebox(0,0)[lb]{\smash{\SetFigFont{10}{12.0}{rm}$d$}}}
\put(3994,-567){\makebox(0,0)[lb]{\smash{\SetFigFont{10}{12.0}{rm}$c$}}}
\put(3994,-2648){\makebox(0,0)[lb]{\smash{\SetFigFont{10}{12.0}{rm}$a$}}}
\put(5569,-2648){\makebox(0,0)[lb]{\smash{\SetFigFont{10}{12.0}{rm}$b$}}}
\put(4838,-1523){\oval(112,112)}
\end{picture}
\label{eq:X}
\end{equation}
The commutation condition \eqref{eq:Isingintegrability} is then equivalent to the existence of a new coupling $M$ such that
\begin{equation}
X(a\, b\, c\, d;\, x_2^\pm,\, x_1^\pm)e^{M(bd)} = e^{M(ac)}X(a\, b\, c\, d;\, x_2^\pm,\, x_1^\pm) \ ,
\label{eq:XM}
\end{equation}
where $M(ab) =\pm M$, depending on whether the spins $a$ and $b$ have the same or opposite orientation. 
Equation \eqref{eq:XM} is the standard Yang-Baxter equation for the Ising model~\cite{Baxter}.

On the other hand, once we write out the matrix products in \eqref{eq:Isingintegrability} using the transfer matrices 
\eqref{eq:V} and \eqref{eq:W}, it is immediate to see that condition \eqref{eq:Isingintegrability} is satisfied iff
\begin{equation}
x_1^+ + \frac{1}{x_1^+} - x_1^- - \frac{1}{x_1^-} = x_2^+ + \frac{1}{x_2^+} - x_2^- - \frac{1}{x_2^-} \ .
\label{eq:integrability}
\end{equation}
Denoting the common value on the LHS and the RHS of \eqref{eq:integrability} 
by $\frac{4i}{g}$ (and allowing $g$ to be an arbitrary complex variable), we see that the Yang-Baxter equations 
for the Ising model are equivalent to \eqref{eq:closure}. In appendix \ref{app:YBequivclosure} we will show that 
the Yang-Baxter equations of the $su(2|2)$ spin chain S-matrix are equivalent to a set of conditions 
of the form \eqref{eq:integrability}. 
\footnote{In earlier works (such as \cite{BeisertS} and \cite{BeisertS2}) it was shown that the closure 
\eqref{eq:closure} implied Yang-Baxter, but complete equivalence was, to our knowledge, never established. 
Notice also that \eqref{eq:closure} is equivalent to the Yang-Baxter equation for the $su(2|2)$ S-matrix. 
This is not the case for the $su(1|2)$ S-matrix \cite{longrange}, which satisfies Yang-Baxter automatically 
for any values of $x^{\pm}$.}

For the Ising model, condition \eqref{eq:integrability} is naturally expressed in terms of $K$ and $L$. Using 
the map \eqref{eq:Isingmap}, 
\begin{equation}
x^+ + \frac{1}{x^+} - x^- - \frac{1}{x^-} = -4\sinh 2K \sinh 2L \ .
\label{eq:xpxmKL}
\end{equation}
Condition \eqref{eq:integrability} thus implies that the expression $\sinh 2K_i \sinh 2L_i$ is the same 
for all $i$. This defines the elliptic modulus $k$ of the Ising model, 
\begin{equation}
\sinh 2K \sinh 2L =\frac{1}{k} \ .
\label{eq:closureKL}
\end{equation}
Combining \eqref{eq:closure} and \eqref{eq:closureKL} we get 
\begin{equation}
k = i \, g \ ,
\label{eq:kgamma}
\end{equation}
relating $k$ and the 't Hooft coupling constant. Notice that this elliptic modulus is the same as 
the one used in \cite{BeisertS2} to parameterize \eqref{eq:closure} in terms of a rapidity $z$.

Let us now consider the partition function of the two-row Ising model in terms of $x^\pm$. 
From \eqref{eq:Z}, taking $m=2$, and using the representation given by \eqref{eq:V} and 
\eqref{eq:W}, one finds that
\begin{equation}
Z = \text{Tr} (VW) = 2\left[ 4 + \left( x^- + \frac{1}{x^-} \right)^2 \right] \ .
\label{eq:Zxm}
\end{equation}
This also has a simple expression in terms of the long-range spin chain variable $u$. Using \eqref{eq:u} 
and \eqref{eq:closure}, $u$ can be rewritten as
\begin{equation}
u = x^- + \frac{1}{x^-} + \frac{2i}{g} \ .
\label{eq:u2}
\end{equation}
The above identification of the Ising elliptic parameter $k=(\sinh 2K \sinh 2L)^{-1}$ with $ig$ 
allows us to rewrite the partition function as
\begin{equation}
Z =  2\left[ 4 + \left( u - \frac{2i}{g} \right)^2 \right] \ .
\label{eq:Zu}
\end{equation}
The spin chain variable $u$ is thus the variable parameterizing the partition function. 


\section{The correlation length and Kramers-Wannier}

\label{sec:corr}

\no
In this section we will reinterpret the correlation length for the Ising model in 
terms of gauge theory variables. Using the Ising model formulae \cite{Baxter}, and the expression
\begin{equation}
u = 2 \cosh 2K \cosh 2L \ ,
\label{eq:uKL}
\end{equation}
obtained by inserting \eqref{eq:Isingmap} into the long-range spin chain variable $u$, \eqref{eq:u}, 
the correlation length is given, for real and positive $k$, as \footnote{Strictly speaking, this formula is obtained in the 
thermodynamic limit. However, the qualitative conclusions we derive from it should be valid in general.}
\begin{equation}
\xi^{-1} = \ln \left( \frac{u/2 + \frac{| 1-k |}{k}}{u/2 - \frac{| 1-k |}{k}} \right) \ .
\label{eq:corr}
\end{equation}
It is easy to check that, for real, positive $k$, this expression is invariant under Kramers-Wannier 
duality,
\be
k \rightarrow 1/k, \quad u \rightarrow k u \ . 
\label{eq:KW}
\ee
Notice also that at the self-dual point $k=1$ the correlation length becomes infinity, indicating the 
existence of a critical point, whose meaning will be discussed in section \ref{sec:critical}. However, in 
our case $k$ is taken as $ig $ which, for real couplings, 
is obviously not a real, positive number, having as a consequence that \eqref{eq:corr} is no longer invariant 
under \eqref{eq:KW}. We should however bear in mind that \eqref{eq:Isingmap} does not define a mapping of the spin 
chain variables to the ordinary Ising model, but to an Ising model analytically continued in $K$ and the elliptic 
modulus $k$. The expression \eqref{eq:corr} for the correlation length, although correct on the positive, 
real $k$-axis, should thus be analytically continued to the entire plane. The most natural way to define the analytic 
extension is to impose invariance under \eqref{eq:KW}, because Kramers-Wannier is a symmetry of the partition function, 
and should therefore be present whether $K$, $L$ and $k$ are real or complex. Thus for general $k$, we will 
define $\xi$ by
\begin{equation}
\xi^{-1} = \ln \left( \frac{u/2 + \frac{\sqrt{ (1-k)^2 }}{k}}{u/2 - \frac{\sqrt{ (1-k)^2 }}{k}} \right) \ ,
\label{eq:corr2}
\end{equation}
which, for a suitably chosen branch of the square roots, obviously coincides with the previous expression 
when $k$ is real and positive, is invariant under Kramers-Wannier duality, and still exhibits a critical point at $k=1$. 

Our main reason for studying the correlation length is the limit which one obtains for small momenta and 
fixed coupling. Since $u = 2\cosh 2K \cosh 2L$, using \eqref{eq:closureKL} and \eqref{eq:uKL} we can write 
\be
u = 2\cosh 2K \sqrt{1+\frac{1}{k^2\sinh^2 2K}} \ .
\ee 
Then, when $|K| \ll |k|^{-1}$ we get
\begin{equation}
u \sim \pm\frac{2 \cosh 2K}{k \sinh 2K} = \mp \frac{2 \cos \frac{p}{2}}{g \sin \frac{p}{2}} \ , 
\quad \text{when } \:\: |p| \ll |g |^{-1} \ .
\label{eq:uKzero}
\end{equation} 
Inserted into \eqref{eq:corr2}, we then get
\begin{equation}
\xi ^{-1} \sim \ln \left( \frac{ \cos \frac{p}{2} \pm i (1-ig)\sin \frac{p}{2} }{\cos \frac{p}{2} 
\mp i (1-ig)\sin \frac{p}{2} } \right) \rightarrow \pm i ( 1 - i\,  g )p \ , \quad \text{ when } p \rightarrow 0 \ .
\label{eq:corrKzero}
\end{equation}
From the previous expression we see that for generic and finite coupling the correlation length becomes 
infinity when the magnon momentum goes to zero. Moreover the real part of the correlation length in the limit 
of strong 't Hooft coupling becomes exactly the string momentum, $p_{\hbox{\tiny{string}}}= g p$ \cite{zaremboworldsheet}, 
while in the limit of small 't Hooft coupling the correlation length is completely determined by the magnon 
momentum. An amusing formal representation of the correlation length in the limit of small magnon momentum 
is thus $\xi^{-1} \sim (p_{\hbox{\tiny{string}}} +i p_{\hbox{\tiny{chain}}})$. 
After our previous discussion on the role of the magnon momentum as parameterizing the deformation of the 
Hopf algebra we observe that the algebra becomes classical precisely when the correlation length becomes infinity, 
{\em i.e.} at the critical points. As we will discuss in the next subsection the critical behaviour at $p=0$, 
and for generic finite coupling, appears because in this limit the model becomes effectively one-dimensional.

The above result is valid for $|K| \ll |k|^{-1}$, but not if $g \rightarrow \infty$ faster than $p \rightarrow 0$. 
This is for instance the case in the near-flat limit considered in \cite{MS}, where a non-trivial S-matrix is obtained. 
This phenomenon can also be understood from the Ising model point of view. In fact if 
$g \rightarrow \infty$ faster that $p \rightarrow 0$, one obtains $u \sim 2\cosh 2K = 2\cos \frac{p}{2}$, and
\be
\xi^{-1} \rightarrow \ln \left( \frac{\cos \frac{p}{2} + 1}{\cos \frac{p}{2} - 1} \right)\rightarrow \infty \ .
\label{eq:corrKinfty}
\ee
We thus see that in the limit $p \rightarrow 0$ the correlation length vanishes.
In Ising model terms this case, where both $K$ and $L$ 
go to zero, is the high-temperature limit. Using \eqref{eq:CL}, we see that the high-temperature regime in the 
Ising model corresponds in the spin chain to the $p \rightarrow 0$ and $C \rightarrow \infty$ regime. 
It is easy to see that in the high-temperature limit the S-matrix might not become trivial when 
$p \rightarrow 0$, although there are also ways of taking this limit which produce a trivial S-matrix. 
These issues will be discussed in the next subsection.
 
For completeness let us also note that a similar situation to the $K \rightarrow 0$ regime will arise when 
$K \rightarrow \infty$. This is due to the symmetry between $K$ and $L$ in the Ising model, since when 
$K \rightarrow \infty$ with $k$ generic the $L$-coupling vanishes, $L \rightarrow 0$, 
and we will once again obtain an infinite 
correlation length (again with an exception, appearing now at weak-coupling, and corresponding to the 
low-temperature limit). This critical behaviour is however 
not observed on the spin chain side, since the latter is defined for real momentum $p$, and $K \rightarrow \infty$ would 
correspond to $p \rightarrow - i\infty$. It is interesting to observe how the analytical continuation in $K$ has broken the 
original symmetry of the Ising model. This is fortunate for us since \eqref{eq:Isingmap} implies that an 
interchange of $K$ and $L$ inverts $x^+$, which is not a symmetry of the spin chain. It also implies that the 
high-temperature limit is observed from the spin chain, but not the low-temperature one.


\subsection{The high-temperature limit and triviality of the S-matrix}

\label{sec:hightemp}

In the previous section, we found that the Ising high-temperature limit $K,\, L \rightarrow 0$, 
had a correlation length $\xi$ of zero, despite that one expects the S-matrix to become trivial for 
small momenta. In this section we will show that it is possible to obtain a non-trivial S-matrix in 
this limit and that it is the only limit which can be non-trivial for $K \rightarrow 0$.
Let us start by showing this last statement. If $K \rightarrow 0$ and we are outside the high-temperature 
limit either $L$ stays finite, or $L \rightarrow \infty$. In the first case, corresponding to the plane-wave 
limit, we can safely take $K$ to zero and set $x^+ = x^-$, obtaining a trivial S-matrix. 
The second case, $L \rightarrow \infty$, arises when $g$ stays finite, or goes to infinity slower than 
$p \rightarrow 0$. We thus see from \eqref{eq:closureKL} that 
$e^{-2L} \sim k\, \sinh 2K$ so that $x^\pm \sim -g \sin \frac{p}{2}e^{\pm ip/2}$. Inserting this 
into the S-matrix of \cite{BeisertS2} one finds that it becomes trivial when $p_1,\, p_2 \rightarrow 0$.

Consider now the high-temperature limit with $K_1 = tK_1'$, $K_2 = tK_2'$ and $k = \delta /t^2$, 
where $K_1'$, $K_2'$ and $\delta$ are 
fixed and where we will let $t \rightarrow 0$. This is the near-flat limit, because 
$p\sqrt{g} \sim p\sqrt[4]{\lambda}$ remains constant \cite{AFS,MS}. Using \eqref{eq:closureKL}, the $L$-couplings can 
be expressed in terms of $t$, $\delta$ and the $K$-couplings. Inserting this into the S-matrix 
it is not difficult to check that, 
by adjusting $K_1'$, $K_2'$ and $\delta$, we can make a given matrix element take any value we like.

Moving on to a more general case, let $K_1 = tK_1'$ and $K_2 = tK_2'$, as above, but where, in the limit 
$t \rightarrow 0$, the dominating contribution to $k$ is of the form $k = \delta /t^\alpha$, for some 
exponent $\alpha$. In order to be in the high-temperature regime we must have $\alpha > 1$ (note that 
the near-flat limit corresponds to $\alpha = 2$). Using \eqref{eq:closureKL} we see that, when $t \rightarrow 0$,
\be
2tK'\cdot 2L = \frac{t^\alpha}{\delta} \; \Rightarrow \; L = \frac{t^{\alpha -1}}{4\delta K'} \ ,
\ee 
so that
\be
x^{\pm} = e^{-2L}e^{\pm 2K} 
\sim 1 \pm 2K't - \frac{t^{\alpha -1}}{2\delta K'} \ .
\ee
Inserting this expression into the S-matrix shows that it is also non-trivial for  $\alpha > 2$. Some matrix 
elements can be chosen arbitrarily, by adjusting $K_1'$, $K_2'$ and $\delta$, while some take fixed, 
constant values. On the other hand, for $1 < \alpha < 2$, the S-matrix becomes trivial once again, 
despite being obtained in a high-temperature limit.

Summing up, we have the following results for the limit $K_1 = tK_1'$, $K_2 = tK_2'$ and $k = \delta /t^\alpha$:
\begin{align*}
\alpha \leq 1  &\Rightarrow \text{Trivial} \\
1 < \alpha < 2 &\Rightarrow \text{Trivial, despite being high-temperature} \\
\alpha = 2 &\Rightarrow \text{Non-trivial, near-flat limit} \\
2 < \alpha  &\Rightarrow \text{Non-trivial .}
\end{align*}
Among these cases the near-flat limit stands out. Besides corresponding to the first non-trivial $\alpha$, 
it is the only small-momentum limit where the value for the S-matrix element $A$ (in the notation 
of \cite{BeisertS2}), which corresponds to the process $\phi \, \phi \rightarrow \phi \, \phi$, describing 
scattering of bosons of the same type, can be adjusted to an arbitrary value. In contrast, when 
$\alpha < 2$ we get  $A = 1$, and when $\alpha > 2$, $A= -1$. 
Also, it is only for $\alpha = 2$ that the S-matrix is sensitive to the value of 
$\delta$. For $\alpha > 2$ all terms containing $\delta$ fall out. However 
when $\alpha > 2$ the limit has also a fascinating property: the matrix elements are 
such that there is virtually no difference between fermions and bosons. \footnote{The S-matrix 
of \cite{BeisertS2} has the additional parameters $\gamma _i$ and $\alpha _B$ which must be, 
in order for the representations to be unitary, equal to, respectively, 
$\sqrt{x_i^+ - x_i^-}$ and 1, up to some phase factors. If we choose to 
include no additional phases the only difference between fermions and bosons 
is a minus sign in a single matrix element.}
  
These results show that an infinite correlation length guarantees triviality in the small-momentum limit, 
but that there are also cases in which the S-matrix is trivial, despite $\xi$ being finite. These are 
the plane-wave limit, where $\alpha = 1$ and $L$ remains finite, so that $u \sim 2\cosh 2L$, and
\be
\xi^{-1} = \ln \left( \frac{\cosh 2L + 1}{\cosh 2L - 1} \right) \ ,
\ee
and the high-temperature limit when $1 < \alpha < 2$, where $\xi =0$. When considering the 
correlation length these results seem slightly out of place. However, it might be worthwile to 
study these limits carefully as their triviality might be a consequence of the simplifications 
used in constructing the theory, notably the assumption of asymptotically long spin chains. 


\subsection{Critical points and the one-dimensional Ising model}

\label{sec:critical}

\no
We will now discuss the meaning of the critical points found in section \ref{sec:corr}. 
The ordinary Ising model critical point $k=1$, has a natural interpretation as a branch point of 
the magnon dispersion relation \footnote{We thank J. Minahan for pointing this out to us.}
\be
E = \sqrt{1 + g^2 \sin ^2 \frac{p}{2}} \ .
\ee
We see that when the sine takes its maximum value, the argument inside the square-root becomes 
zero precisely when $g = \pm i$ or, equivalently, $k=\pm 1$. This critical behaviour can thus be found 
at the boundary of the domain of convergence of the planar gauge theory. In this sense the Ising 
phase transition can be reflecting the regime where the number of planar diagrams becomes dense 
(see \cite{Polyakov} for a recent discussion). 
  
The other critical points, obtained for small momentum, are not standard two-dimensional Ising model 
critical points, but they can be understood in terms of the one-dimensional Ising model. As we have seen, 
the critical behaviour arises when $K \rightarrow \infty$, 
or $L \rightarrow \infty$. This implies that, in general, along one of the directions the spins will 
always have the same orientation, as it would cost an infinite amount of energy to let two adjacent spins 
have opposite orientations. This means that we effectively obtain a one-dimensional model, and the partition 
function for the two-dimensional Ising model becomes the one-dimensional partition function, up to 
an infinite constant. It is well known that the one-dimensional Ising model has a critical point at 
zero temperature, and it is precisely this point that is obtained.

For a one-dimensional Ising model, the correlation length is given, in terms of the coupling $K^{(1D)}$, as
\be
\xi _{(1D)}^{-1} = \ln \left( \frac{\cosh K^{(1D)}}{\sinh K^{(1D)}}  \right) 
= \ln \left( \frac{e^{2K^{(1D)}} + 1 }{e^{2K^{(1D)}} - 1}\right).
\label{eq:corr1D}
\ee
In the case $K \rightarrow 0$, studied above, the correlation length was given by \eqref{eq:corrKzero}, 
which written in terms of $K$ and $k$ is
\be
\xi ^{-1} = \ln \left( \frac{ \cosh 2K \pm (1-k)\sinh 2K }{\cosh 2K \mp (1-k)\sinh 2K } \right).
\label{eq:corr2Dzero}
\ee
Combining \eqref{eq:corr1D} and \eqref{eq:corr2Dzero} allows us to, for a given $k$, relate $K$ and $K^{(1D)}$. 
In general, however, we can see that $K \rightarrow 0$ corresponds to $K ^{(1D)} \rightarrow \infty$, which is the 
low-temperature limit of the one-dimensional model, where its only critical point can be found.
  
The relationship is especially simple in the limit of zero coupling, which for $K$ fixed also gives 
$L \rightarrow \infty$. In this regime it is easy to check that $\xi$ is given by \eqref{eq:corr2Dzero}, with $k=0$, 
independently of the value of $K$. Taking the plus sign, we then have the identification
\be
\frac{\cosh 2K}{\sinh 2K} = e^{2K^{(1D)}_{k = 0}}.
\label{eq:KKzero}
\ee
At infinite coupling, for fixed $K$, we do not have $L \rightarrow \infty$, but rather $L \rightarrow 0$. This 
can also be interpreted as a one-dimensional model, because now the horizontal rows decouple from each other, 
and we get indeed a set of one-dimensional models. The correlation length is now as in \eqref{eq:corrKinfty}, and
\be
\cosh 2K = e^{2K^{(1D)}_{g = \infty}},
\label{eq:KKinfty}
\ee
and the critical point now corresponds to $K \rightarrow \infty$. In this limit, $e^{2K} \rightarrow e^{2K^{(1D)}_{g = \infty}}$. 


\subsection{Kramers-Wannier duality and the long-range Bethe ansatz}

\no
Before we conclude this section, let us study the Kramers-Wannier duality transformation 
a little more closely. We will first focus on the transformation of the long-range variable $u$. 
From \eqref{eq:KW}, we see that $u \rightarrow k\, u$ 
under the duality. The inverse is also true: if we impose the one-loop Heisenberg-model result, which rescaled 
in order to match our conventions 
takes the form
\be
u _{\text{one-loop}}(p)=\frac{2}{g}\cot \frac{p}{2},
\label{eq:uheisrescaled}
\ee
then, as is shown in appendix \ref{app:KW}, imposing Kramers-Wannier will give the all-loop result
\be
u(p) = \frac{2}{g}\cot \frac{p}{2}\sqrt{1 + g^2 \sin ^2 \frac{p}{2}},
\label{eq:uall-looprescaled}
\ee
at least as the minimal solution.

Let us now discuss in some detail the Kramers-Wannier duality as a modular transformation. 
The map into the Ising model that we have suggested provides indeed a natural way to analitically extend 
${\cal N}=4$ Yang-Mills to arbitrary complex coupling. When we consider the $k$-plane, with 
$k=ig$, the physical region with positive coupling is just the positive imaginary axis. The 
negative imaginary axis corresponds thus to the analytic extension $\sqrt{\lambda} \rightarrow 
- \sqrt{\lambda}$, that could be holographically interpreted as some sort of continuation 
into de Sitter space \cite{Polyakov}. The Kramers-Wannier duality transformation 
$k \rightarrow 1/k$ maps one region into the other. \footnote{Another possibility is to consider 
BPS bound states \cite{Dorey}. The integrability constant is then replaced by $4in/g$, and the 
Kramers-Wannier transformation is interpreted as a strong/weak-coupling duality 
transforming the 't Hooft coupling into $n^2/g$, and a global change of sign.}
  
We can lift transformations on the elliptic modulus $k$ into $Sl(2,\mathbb{Z})$ modular 
transformations of the underlying elliptic curve, with complex moduli $\tau(k)=i K'(k)/K(k)$. The 
transformation $k \rightarrow ik/k'$, with $k^2+ k'^2 = 1$, corresponds to $\tau \rightarrow 
\tau +1$, and $k \rightarrow k'$ to $\tau \rightarrow -1/\tau$. The Kramers-Wannier duality 
transformation can be represented as the composition of $k \rightarrow -k$, and 
$g \rightarrow 1/g$. The first one can be recovered by performing twice the 
transformation $k \rightarrow ik/k'$, and therefore can be interpreted as the 
modular transformation $\tau \rightarrow \tau + 2$. The second is a standard strong/weak-coupling 
transformation, for the 't Hooft coupling. Keeping in 
mind the last fact, that the AdS/CFT correspondence is a duality on the 't Hooft coupling, it is 
interesting to wonder how much of this strong/weak-coupling duality is captured by the 
Kramers-Wannier transformation. In fact the $g \rightarrow 1/g$ transformation can be 
approached from a different point of view \cite{BeisertS2}. From the modular transformations 
of $\tau$ one gets
\be
\tau \left( - ik/k' \right) = - \frac {1}{\tau \left( 1/k' \right)} \ ,
\ee
that for $k=ig$ is equivalent to the transformation $g \rightarrow 1/g$. Notice also that 
the self-dual point for this transformation, $g=1$, does not correspond to a special 
point from the Ising model point of view. However at this point Kramers-Wannier corresponds to 
just the change of sign $k \rightarrow -k$. From the Ising model point of view 
the special point corresponds a complex value for the coupling, $g=i$, which is the 
place where the Ising phase transition takes place, and where as we have discussed above 
we reach the radius of converge of the planar series.


\section{Conclusions}

\no
The present understanding of the map from ${\cal N}=4$ dynamics to integrable spin chains 
is based on two key ingredients. One is the classical algebra of symmetries. The other is the set of 
kinematical relations defining the $x^\pm$ variables. Both ingredients become entangled in the underlying 
Hopf algebra structure. What we see within the description of the long-range spin chain in terms of the Ising model 
is that the kinematical ingredients encoded in the central Hopf subalgebra are just defining a two-dimensional 
Ising model, and that the integrability of the spin chain is completely captured by the underlying Ising model 
Yang-Baxter equation. 
The results of this note can be summarized in the following table of correspondences between the spin-chain 
magnon dynamics and the two-dimensional Ising 
\begin{table}[h]
\begin{center}
\vspace{6 mm}
\hspace{30 pt}\begin{tabular}{|lcl|}
\hline
Magnon variables $x^{\pm}$ & $\Leftrightarrow$ & Transfer matrix elements \\
$x^+ + \frac{1}{x^+} - x^- - \frac{1}{x^-} =$ {invariant} & $\Leftrightarrow$ & Yang-Baxter condition \\
't Hooft coupling & $\Leftrightarrow$ & Elliptic parameter \\
Departure from triviality & $\Leftrightarrow$ & Inverse correlation length \\
Planar convergence radius & $\Leftrightarrow$ & Ising critical point \\
\hline
\end{tabular}
\caption{Equivalence of ${\cal N}=4$ Yang-Mills variables to Ising model parameters.}
\end{center}
\end{table}
model. Probably the most interesting part of the correspondence is 
the one to one relation between the Yang-Baxter equation for the $su(2|2)$ spin chain magnon S-matrix and the Yang-Baxter 
equation for the two-dimensional Ising model. An obvious consequence of this correspondence is that both the 
${\cal N}=4$ Yang-Mills spin chain and the Ising model share the same elliptic curve. 
Another interesting consequence of this equivalence is the 
relation between the magnon momentum and the Ising model correlation length. The magnon 
momentum operator is the additional piece that must be added to the symmetry algebra $su(2|2)$ in 
order to get a Hopf algebra determining a non-trivial S-matrix. On the other hand, the correlation length 
measures the departure from critical behaviour. We have described how critical 
behaviours, corresponding to infinite correlation lengths, are in correspondence with those zero-momentum 
limits of the spin chain where the S-matrix becomes trivial. There is however a piece in the scattering matrix that 
we have not considered at all in the present work. This is the global dressing phase factor, responsible 
for the interpolation from the strong to the weak-coupling regime. Hopefully the equivalence to the Ising 
model will also provide some light on the general structure of this dressing factor. 
  

\vspace{5mm}
\centerline{\bf Acknowledgments}

\no
We are grateful to N. Dorey, F. Ferrari, J. Minahan and R. Roiban for comments and discussions. 
C. G. is also grateful to the Isaac Newton Institute for Mathematical Sciences for 
hospitality while this work was being completed. The work of J. G. is supported by 
a Spanish FPU grant and by a European fellowship through MRTN-CT-2004-005104. 
This work is also partially supported by the Spanish DGI under contracts 
FPA2003-02877 and FPA2003-04597 and by the CAM project HEPHACOS P-ESP-00346. 


\appendix

\section{Yang-Baxter and the closure condition}

\label{app:YBequivclosure}

\no
In this appendix we will demonstrate that the Yang-Baxter equations obtained from the S-matrix of 
\cite{BeisertS2}, derived by imposing invariance under maximally centrally extended-$su(2|2)$, are equivalent 
to conditions of the form \eqref{eq:Isingintegrability}, or equivalently, to the closure \eqref{eq:closure}. 
From the algebraic construction itself \cite{BeisertS2}, or by direct computation using computer software, it follows 
that the closure condition \eqref{eq:closure} implies the Yang-Baxter equations of the S-matrix. The question that 
then arises is if the closure is necessary for Yang-Baxter to hold. The answer is not as obvious as it might seem. 
In \cite{longrange}, the S-matrix for the $su(1|2)$-sector of the theory was constructed, also using the spectral 
variables $x^+$ and $x^-$ (albeit scaled differently with respect to the convention used here), and it was found 
that the Yang-Baxter equation was satisfied without having to impose any relation between $x^+$ and $x^-$. 

We will now settle the issue of the equivalence of Yang-Baxter and the closure. Fortunately, one of the 
Yang-Baxter equations takes an exceptionally simple form, solving the problem for us. Using the matrix elements 
and the notation of \cite{BeisertS2}, the equation corresponding to the process 
$|\phi^1 \psi^1 \phi ^2 \rangle \rightarrow |\psi^2 \psi^1 \psi ^1 \rangle $ is 
\footnote{Here, we have set the marker variable $\xi _k =1$. This is permitted since all the $\xi _k$ cancel 
from the Yang-Baxter equations, and we thus get the same result as with the Hopf algebra compatible 
value $\xi _k = \sqrt{\frac{x_k^+}{x_k^-}}$.} 
\be
\frac{1}{2}H_{12}L_{13}C_{23} + \frac{1}{4}G_{12}C_{13}(D_{23}-E_{23})\frac{x_2^-}{x_2^+} = \frac{1}{2}L_{23}C_{13}D_{12} \ ,
\ee
which, after plugging in the expressions for the matrix elements and simplifying, becomes
\be
x^+_2 + \frac{1}{x^+_2} - x^-_2 - \frac{1}{x^-_2}  = x^+_3 + \frac{1}{x^+_3} - x^-_3 - \frac{1}{x^-_3} \ .
\label{eq:xclosure23}
\ee
Furthermore, the equation for the process $|\phi^1 \phi^2 \psi ^1 \rangle \rightarrow |\psi^1 \psi^1 \psi ^2 \rangle $ is 
\be
\frac{1}{2}C_{12}(D_{13}-E_{13})D_{23} = \frac{1}{4}H_{23}C_{13}(D_{12}-E_{12}) + \frac{1}{2}G_{23}G_{13}C_{12}\frac{x_3^-}{x_3^+} \ ,
\ee
which becomes
\begin{align}
& \Big( x^-_2 + \frac{1}{x^-_2} - x^-_1 - \frac{1}{x^-_1})(x^+_3 + \frac{1}{x^+_3} - x^+_1 - \frac{1}{x^+_1} \Big) = \nonumber \\
=& \Big( x^+_2 + \frac{1}{x^+_2} - x^+_1 - \frac{1}{x^+_1})(x^-_3 + \frac{1}{x^-_3} - x^-_1 - \frac{1}{x^-_1} \Big) \ . \label{eq:YB13}
\end{align}
Using \eqref{eq:xclosure23}, this can be rewritten as
\be
x^+_1 + \frac{1}{x^+_1} - x^-_1 - \frac{1}{x^-_1}  = x^+_3 + \frac{1}{x^+_3} - x^-_3 - \frac{1}{x^-_3} \ .
\label{eq:xclosure13}
\ee
We have thus shown that the Yang-Baxter equation implies that the quantity 
$x^+_i + \frac{1}{x^+_i} - x^-_i - \frac{1}{x^-_i}$, for $i=1,\, 2,\, 3$, is equal to a common value, 
\footnote{It should be noted that there is a subtlety in this calculation. Some of the matrix elements, 
presented in \cite{BeisertS2}, were simplified using \eqref{eq:closure}. In fact, the S-matrix of \cite{BeisertS2} 
does not satisfy \eqref{eq:Smatrixconstruction}, if one does not impose \eqref{eq:closure}. This means that the equations 
that we have just derived could just as well be an artifact of this simplification, and that a non-simplified S-matrix would 
yield trivially satisfied Yang-Baxter equations. Fortunately, this is not the case. We have re-derived the matrix elements, 
as determined by equation \eqref{eq:Smatrixconstruction}, but without using \eqref{eq:closure} to simplify them, and 
checked that the closure is indeed necessary for Yang-Baxter to be satisfied.} which we may call $4i/g$. 
Therefore the Yang-Baxter equation implies that
\be
x^+_j + \frac{1}{x^+_j} - x^-_j - \frac{1}{x^-_j} = \frac{4i}{g} \ , \quad j=1,\, 2,\, 3 \ .
\label{eq:YBimplyclosure}
\ee
This $g$ does not have to be constant, though, and the Yang-Baxter equations are satisfied no matter how 
complicated $g$ may be. If we want to interpret $g$ as a coupling constant, we are forced to draw 
the conclusion that the physically admissable solutions of the Yang-Baxter equations are only a small part 
of the entire set of solutions.


\section{Proof  that Kramers-Wannier duality determines $u$}
\label{app:KW}

\no 
In this appendix we will show that imposing Kramers-Wannier duality \eqref{eq:KW}, and the one-loop result
\be
u _{\text{one-loop}}(p)=\frac{2}{g}\cot \frac{p}{2} \ ,
\label{eq:uheisrescaled2}
\ee
gives the all-loop result
\be
u(p) = \frac{2}{g}\cot \frac{p}{2}\sqrt{1 + g^2 \sin ^2 \frac{p}{2}} \ .
\label{eq:uall-looprescaled2}
\ee
To show this, we must use the transformation properties of $K$ under 
Kramers-Wannier duality. In fact, there are two ways to define $K^\ast$, the dual of $K$, compatible with 
\eqref{eq:KW}, $\sinh 2K^\ast = \frac{1}{\sinh 2K}$, and $\sinh 2K^\ast = k \cdot \sinh 2K$. Here we will 
show the statement for  
\be
\sinh 2K^\ast = k \cdot \sinh 2K \Leftrightarrow \sin \frac{p^\ast}{2} = ig \sin \frac{p}{2} \ ,
\label{eq:pdual}
\ee
but the other option produces the same result. 
Let us now set $x \equiv \sin \frac{p}{2}$. From \eqref{eq:pdual}, Kramers-Wannier is then 
\begin{align}
x^\ast &= kx \ , \nonumber \\
k^\ast &= k \ , \label{eq:KW2} \\ 
u ^\ast &= h(k) u \ , \nonumber
\end{align}
where we, for the moment, leave a function $h(k)$ free. The one-loop 
expression \eqref{eq:uheisrescaled2} implies that in general $u$ takes the form
\be
u = \frac{2i}{k}\frac{\sqrt{1 - x^2}}{x}\cdot f(x,\, k) \ ,
\label{eq:uform}
\ee
where $f(x,\, 0)=0$ for all possible values of $x$. Now, applying \eqref{eq:KW2} to \eqref{eq:uform} and 
using the transformation properties of $u$ gives
\be
\frac{2i h(k)}{k}\frac{\sqrt{1 - x^2}}{x}\cdot f(x,\, k) = \frac{\sqrt{1 - k^2 x^2}}{x} \cdot f(kx,\, k^{-1}) \ ,
\ee
which leads to the following equations for the function $f$,
\begin{align}
\frac{f(kx,\, k^{-1})}{f(x,\, k)} &= \frac{k\sqrt{1 - x^2}}{h(k)\sqrt{1 - k^2 x^2}} \ , \nonumber \\
f(x,\,0) &= 1 \ . \label{eq:eqforf}
\end{align}
If we now define a new function $m$ by
\be
m(x,\, y) \equiv \frac{f(x.\, y)}{\sqrt{1 - (xy)^2}} \ ,
\label{eq:defg}
\ee
equation \eqref{eq:eqforf} becomes
\begin{align}
\frac{m(kx,\, k^{-1})}{m(x,\, k)} &= \frac{k}{h(k)} \ ,\nonumber \\
m(x,\,0) &= 1 \ . \label{eq:eqform}
\end{align}
The dependence in $x$ must cancel from the left-hand side of \eqref{eq:eqform}. There are two ways 
in which this can happen, the first being the direct cancellation of the entries in the first argument 
of $m$, and the second being if $m$ has Kramers-Wannier invariant factors, such as $(1+ \sqrt{k}x)$. For 
the moment, let us suppose that $m$ does not have any Kramers-Wannier invariant factor.

In order for the arguments in the first entry of $m$ to cancel, it must then take the form
\be
m(x,\, y) = x^\eta g(y) \ ,
\ee
and \eqref{eq:eqform} then implies that $\eta = 0$. Thus $m$ is a momentum-independent function. Retracing our steps, we find
\be
u = \frac{2i}{k}\frac{\sqrt{1 - x^2}}{x}\sqrt{1 - (xk)^2} m(k) = 
\frac{2}{g}\cot \frac{p}{2}\sqrt{1 + g ^2 \sin^2 \frac{p}{2}} m(ig) \ .
\ee
Inserting the Kramers-Wannier invariant factors that we, optionally, could have in \eqref{eq:eqform}, we arrive 
at the complete solution
\be
u = \frac{2}{g}\cot \frac{p}{2}\sqrt{1 + g ^2 \sin^2 \frac{p}{2}} m(ig) q\left(\sin \frac{p}{2},\, g\right) \ ,
\ee
where $q$ is Kramers-Wannier invariant and where $\lim _{g \rightarrow 0} m(ig) q\left(\sin \frac{p}{2},\, g\right) = 1$. 
\footnote{This condition is more constraining than it first might seem. For example, $\tilde{u} = \sqrt{g}u$ is a 
Kramers-Wannier invariant, which however cannot appear in $q$. The reason for this is that by controlling the 
behaviour of $p$ when $g \rightarrow 0$, $\tilde{u}$ can be made to take any value that we like, and $q$ can 
thus impossibly satisfy the weak-coupling condition.} There is no way that the function $q$ can cancel the 
factors $\cot \frac{p}{2}$, or $\sqrt{1 + g ^2 \sin^2 \frac{p}{2}}$, because if it contains one of them, it 
must contain the other, and a correction by $\sqrt{g}$, in order to be invariant. Then, it can impossibly 
have the correct weak-coupling behaviour. In our case, we have $h(k)=k$, \footnote{In fact, the problem has 
solutions for all $h(k)$ satisfying $h(k^{-1})=h(k)^{-1}$.} allowing us to take the minimal solution $m\, q = 1$, 
which produces \eqref{eq:uall-looprescaled2}.




\begin{thebibliography}{99}

\renewcommand{\baselinestretch}{1}
\normalsize

\bibitem{MZ} J.A.~Minahan and K.~Zarembo, {\em The Bethe-ansatz for $\mathcal{N}=4$ super Yang-Mills},
JHEP {\bf 0303} (2003) 013, {\tt hep-th/0212208}.

\bibitem{complete} N.~Beisert, C.~Kristjansen and M.~Staudacher,
{\em The dilatation operator of N = 4 super Yang-Mills theory},
Nucl.\ Phys.\  B {\bf 664}, 131 (2003), {\tt hep-th/0303060}.
N.~Beisert and M.~Staudacher,
{\em The ${\cal N} = 4$ SYM integrable super spin chain},
Nucl.\ Phys.\  B {\bf 670}, 439 (2003), 
{\tt hep-th/0307042}.

\bibitem{BDS} N.~Beisert, V.~Dippel and M.~Staudacher,
{\em A novel long range spin chain and planar ${\cal N} = 4$ super Yang-Mills},
JHEP {\bf 0407} (2004) 075, {\tt hep-th/0405001}.

\bibitem{StaudacherS} M.~Staudacher, {\em The factorized S-matrix of CFT/AdS},
JHEP {\bf 0505} (2005) 054, {\tt hep-th/0412188}.

\bibitem{longrange} N.~Beisert and M.~Staudacher,
{\em Long-range $PSU(2,2|4)$ Bethe ansaetze for gauge theory and strings},
Nucl.\ Phys.\  B {\bf 727}, 1 (2005), {\tt hep-th/0504190}.

\bibitem{zaremboworldsheet} T.~Klose, T.~McLoughlin, R.~Roiban and K.~Zarembo,
{\em Worldsheet scattering in $AdS_5 \times S^5$}, JHEP {\bf 0703} (2007) 094, 
{\tt hep-th/0611169}.
T.~Klose, T.~McLoughlin, J.~A.~Minahan and K.~Zarembo,
{\em World-sheet scattering in $AdS_5 \times S^5$ at two loops},
JHEP {\bf 0708} (2007) 051, {\tt arXiv:0704.3891 [hep-th]}.

\bibitem{BeisertS} N.~Beisert, {\em The $su(2|2)$ dynamic $S$-matrix}, {\tt hep-th/0511082}.

\bibitem{BeisertS2} N.~Beisert,
{\em The analytic Bethe ansatz for a chain with centrally extended $su(2|2)$ symmetry},
J.\ Stat.\ Mech.\  {\bf 0701} (2007) P017, {\tt nlin.si/0610017}.

\bibitem{AFS} G.~Arutyunov, S.~Frolov and M.~Staudacher,
{\em Bethe ansatz for quantum strings},
JHEP {\bf 0410} (2004) 016, {\tt hep-th/0406256}.

\bibitem{HL} N.~Beisert and A.~A.~Tseytlin,
{\em On quantum corrections to spinning strings and Bethe equations},
Phys.\ Lett.\  B {\bf 629} (2005) 102, {\tt hep-th/0509084}.
R.~Hern\'andez and E.~L\'opez,
{\em Quantum corrections to the string Bethe ansatz},
JHEP {\bf 0607} (2006) 004, {\tt hep-th/0603204}.
G.~Arutyunov and S.~Frolov, {\em On $AdS_5 \times S^5$ string S-matrix},
Phys.\ Lett.\  B {\bf 639} (2006) 378, {\tt hep-th/0604043}.
L.~Freyhult and C.~Kristjansen,
{\em A universality test of the quantum string Bethe ansatz},
Phys.\ Lett.\  B {\bf 638} (2006) 258, {\tt hep-th/0604069}.

\bibitem{Gromov} N.~Gromov and P.~Vieira,
{\em Constructing the AdS/CFT dressing factor},
Nucl.\ Phys.\  B {\bf 790} (2008) 72, {\tt hep-th/0703266}.
H.~Y.~Chen, N.~Dorey and R.~F.~Lima Matos,
{\em Quantum scattering of giant magnons},
JHEP {\bf 0709} (2007) 106, {\tt arXiv:0707.0668 [hep-th]}.

\bibitem{Janik} R. A.~Janik, {\em The $AdS_5\times S^5$ superstring worldsheet S-matrix and crossing symmetry},
Phys.\ Rev.\  D {\bf 73} (2006) 086006, {\tt hep-th/0603038}.

\bibitem{BHL} N.~Beisert, R.~Hern\'andez and E.~L\'opez,
{\em A crossing-symmetric phase for $AdS_5 \times S^5$ strings},
JHEP {\bf 0611} (2006) 070, {\tt hep-th/0609044}.

\bibitem{BES} N.~Beisert, B.~Eden and M.~Staudacher,
{\em Transcendentality and crossing}, 
J.\ Stat.\ Mech.\  {\bf 0701} (2007) P021
{\tt hep-th/0610251}.

\bibitem{nonpert} C.~G\'omez and R.~Hern\'andez,
{\em Integrability and non-perturbative effects in the AdS/CFT correspondence},
Phys.\ Lett.\  B {\bf 644} (2007) 375, {\tt hep-th/0611014}.

\bibitem{Hopf} C.~G\'omez and R.~Hern\'andez,
{\em The magnon kinematics of the AdS/CFT correspondance}, JHEP {\bf 0611} (2007) 021, 
{\tt hep-th/0608029}.

\bibitem{Plefka} J.~Plefka, F.~Spill and A.~Torrielli,
{\em On the Hopf algebra structure of the AdS/CFT S-matrix}, 
Phys.\ Rev.\ D\ {\bf 74} (2006) 066008, {\tt hep-th/0608038}.

\bibitem{AFZF} G.~Arutyunov, S.~Frolov and M.~Zamaklar,
{\em The Zamolodchikov-Faddeev algebra for $AdS_5 \times S^5$ superstring},
JHEP {\bf 0704} (2007) 002, {\tt hep-th/0612229}.

\bibitem{BeisertYangian} N.~Beisert, {\em The S-Matrix of AdS/CFT and Yangian Symmetry}, 
PoS {\bf SOLVAY} (2006) 002, {\tt arXiv:0704.0400 [nlin.SI]}.

\bibitem{Young} C.~A.~S.~Young,
{\em q-Deformed supersymmetry and dynamic magnon representations},
J.\ Phys.\ A  {\bf 40} (2007) 9165, {\tt arXiv:0704.2069 [hep-th]}.

\bibitem{Lipatov} A.~V.~Kotikov, L.~N.~Lipatov, A.~Rej, M.~Staudacher and V.~N.~Velizhanin,
{\em Dressing and Wrapping}, J.\ Stat.\ Mech.\  {\bf 0710} (2007) P10003, 
{\tt arXiv:0704.3586 [hep-th]}.

\bibitem{Baxter} R.~J.~Baxter, {\em Exactly solved models in statistical 
mechanics}, Academic Press, London (1982).

\bibitem{Kostov} I.~Kostov, D.~Serban and D.~Volin, {\em Strong coupling limit of Bethe ansatz equations}, 
Nucl.\ Phys.\  B {\bf 789} (2008) 413, {\tt hep-th/0703031}.

\bibitem{Martins1} M. J. Martins and P. B. Ramos, {\em The quantum inverse scattering method 
for Hubbard like models}, Nucl. Phys. {\bf B522}, 413 (1998), {\tt solv-int/9712014}.

\bibitem{Martins2} M.~J.~Martins and C.~S.~Melo,
{\em The Bethe ansatz approach for factorizable centrally extended S-matrices},
Nucl.\ Phys.\  B {\bf 785}, 246 (2007)
{\tt hep-th/0703086}.

\bibitem{Hubbard} A.~Rej, D.~Serban and M.~Staudacher,
{\em Planar ${\cal N} = 4$ gauge theory and the Hubbard model}, 
JHEP {\bf 0603}, 018 (2006), {\tt hep-th/0512077}.

\bibitem{MS} J.~M.~Maldacena and I.~Swanson,
{\em Connecting giant magnons to the pp-wave: An interpolating limit of $AdS_5 \times S^5$},
Phys.\ Rev.\  D {\bf 76} (2007) 026002, {\tt hep-th/0612079}.

\bibitem{Polyakov} A.~M.~Polyakov, {\em De Sitter space and eternity},
{\tt arXiv:0709.2899 [hep-th]}.

\bibitem{Dorey} N.~Dorey, D. M.~ Hofman and J.~ Maldacena, {\em On the Singularities of the Magnon S-matrix}, 
Phys.\ Rev.\  D {\bf 76} (2007) 025011, {\tt hep-th/0703104}.

\end{thebibliography}
\end{document}